\documentclass[twocolumn, twocolappendix, times]{aastex631}
\usepackage{txfonts}

\submitjournal{ApJ}
\accepted{30 May 2024}

\begin{document}

\title{Simulating the arrival of the southern substructure in the galaxy cluster Abell 1758}

\shorttitle{Simulating the southern substructure in A1758}
\shortauthors{Machado et al.}

\author[0000-0001-7319-297X]{Rubens E. G. Machado}
\affiliation{Departamento Acad\^emico de F\'isica, Universidade Tecnol\'ogica Federal do Paran\'a, Rua Sete de Setembro 3165, Curitiba, Brazil}
\email{rubensmachado@utfpr.edu.br}

\author[0000-0002-2939-5128]{Ricardo C. Volert}
\affiliation{Departamento Acad\^emico de F\'isica, Universidade Tecnol\'ogica Federal do Paran\'a, Rua Sete de Setembro 3165, Curitiba, Brazil}

\author[0009-0006-9462-1044]{Richards P. Albuquerque}
\affiliation{Departamento Acad\^emico de F\'isica, Universidade Tecnol\'ogica Federal do Paran\'a, Rua Sete de Setembro 3165, Curitiba, Brazil}

\author[0000-0001-6419-8827]{Rog\'erio Monteiro-Oliveira}
\affiliation{Institute of Astronomy and Astrophysics, Academia Sinica, Taipei 10617, Taiwan}

\author[0000-0001-5606-4319]{Gast\~ao B. Lima Neto}
\affiliation{Instituto de Astronomia, Geof\'isica e Ci\^encias Atmosf\'ericas, Universidade de S\~ao Paulo, Rua do Mat\~ao 1226, S\~ao Paulo, SP, Brazil}

\begin{abstract}

{Abell 1758 ($z\sim0.278$) is a galaxy cluster composed of two structures: A1758N and A1758S, separated by $\sim$2.2\,Mpc. The northern cluster is itself a dissociative merging cluster that has already been modelled by dedicated simulations. Recent radio observations revealed the existence of a previously undetected bridge connecting A1758N and A1758S.}
{New simulations are now needed to take into account the presence of A1758S. We wish to evaluate which orbital configuration would be compatible with a bridge between the clusters.}
{Using $N$-body hydrodynamical simulations that build upon the previous model, we explore different scenarios that could have led to the current observed configuration. Five types of orbital approaches were tested: radial, tangential, vertical, post-apocentric, and outgoing.}
{We found that the incoming simulated scenarios are generally consistent with mild enhancements of gas density between the approaching clusters. The mock X-ray images exhibit a detectable bridge in all cases. Compared to measurements of \textit{Chandra} data, the amplitude of the X-ray excess is overestimated by a factor of $\sim$2--3 in the best simulations. The scenario of tangential approach proved to be the one that best matches the properties of the profiles of X-ray surface brightness. The scenarios of radial approach of vertical approach are also marginally compatible.}

\end{abstract}


\section{Introduction}

In merging galaxy clusters, compression of the gas between the two approaching clusters should boost the X-ray emission in that region. Examples of bridges or filaments detected in X-rays between clusters include, among others: A3391 and A3395 \citep{Tittley2001, Biffi2022}, A222 and A223 \citep{Werner2008}, A3556 and A3558 \citep{Ursino2015}, and A399 and A401 \citep{Akamatsu2017}. From the simulation point of view, such enhancements of X-ray surface brightness bridges linking merging clusters were studied by e.g., \citet{Roettiger1997, Ricker2001, Ritchie2002, Donnert17, Chadayammuri22}. These are, in principle, thermal emissions associated with the hot gas, and the increments in X-rays surface brightness are due to increments in gas density.

In radio, it is common for merging clusters to exhibit extended diffuse emission in the form of giant radio relics \citep[e.g.,][]{vanWeeren2019,Monteiro-Oliveira22}, which are typically giant arcs in the periphery of merging clusters, although more complicated radio morphologies are also seen. On the other hand, bridges of radio emission (diffuse synchrotron radiation) connecting pre-merging clusters are a recently discovered entity. At the moment, such radio bridges are known to have been detected in only two pairs or pre-merging clusters: Abell~399/401 \citep{Govoni2019} and Abell~1758N/1758S \citep{Botteon2020}. In the case of Abell~399/401, the radio bridge has a projected length of 3\,Mpc. Theoretically, the mechanisms of magnetic field amplification and particle acceleration are understood to be caused by the probable compression and turbulence generated in the  low-density regions between the clusters \citep{Brunetti2020}.

Abell 1758 (hereafter A1758), at a redshidft of $z\sim0.278$, consists of two major structures: the northern (A1758N) and the southern (A1758S) components, separated by approximately $\sim$2.2\,Mpc. This bimodality has been known since ROSAT observations \citep{Rizza1998}. The northern component is itself bimodal, as indicated by gravitational weak lensing analyses \citep{Dahle2002, Okabe2008, Ragozzine2012, Monteiro2017}. X-ray observations \citep{David2004} combined with optical and multi-wavelength analyses \citep{Boschin2012, Durret2011} further indicated that A1758N must be a post-merger cluster, i.e.~the result of a collision between two clusters. Following the nomenclature of \cite{Monteiro2017}, the two components within A1758N are the NW and NE substructures.

Hydrodynamical $N$-body simulations are often used to model individual merging clusters. These can be used to study different phenomena such as dissociation between dark matter (DM) and gas \citep[e.g.][]{Springel2007, Moura2021, Lourenco2020, Albuquerque2024}, sloshing \citep[e.g.][]{Roediger2011, MachadoLimaNeto2015, Walker2017, Doubrawa2020}, and radio relics \citep[e.g.][]{vanWeeren2011, Lee2020} among many others. In particular, \cite{Ha2018} studied simulations of equatorial shocks, which propagate perpendicularly to the merger axis. \cite{Molnar2013} simulated pre-merger shocks in Abell~1750.

In \cite{Machado2015}, we proposed a tailored simulation model of A1758N, which reproduced several of its observed features, by means of an off-axis collision between two equal-mass clusters. The main constraint was the projected separation of approximately 750\,kpc between the mass peaks of the NW and NE substructures, derived from weak lensing \citep{Okabe2008, Ragozzine2012}. Starting from a total A1758N virial mass of $10^{15}\,{\rm M}_{\odot}$, a mass ratio of $1:1$ was assumed. Furthermore, the peculiar feature of A1758N is that it is a dissociative cluster, meaning that in one of its substructures that gas has been separated from the DM. Namely, the NW substructure coincides with the location of a peak in X-ray emission, while the NE substructure does not. In other words, the NW cluster has retained its gas, while the NE cluster has been mostly stripped of its gas due to the collision. The simulation in \cite{Machado2015} was able to recover, within an acceptable approximation, the overall morphology of the gas at the instant when the separation between NW and NE reached the target value of 750\,kpc.

The simulation model of \cite{Machado2015} was later updated using the more recent weak lensing mass estimates in \cite{Monteiro2017}. Specifically, the new virial masses were $(7.9 \pm 1.6) \times 10^{14}\,{\rm M}_{\odot}$ for NW; and $(5.5 \pm 1.6) \times 10^{14}\,{\rm M}_{\odot}$ for NE. The updated simulation proved to be qualitatively similar to the earlier model. By requiring it to meet the same criteria as before, the best-matching instant was found to be approximately 0.27\,Gyr after the first pericentric passage. That new simulation ensured that the more recent mass estimates remained compatible with the scenario of a DM/gas dissociation induced by the merger. In the present paper, we will take the definitive simulation from \cite{Monteiro2017} to be the fiducial model of A1758N.

The presence of A1758S was disregarded in \cite{Machado2015} and \cite{Monteiro2017}, because at the time there was no evidence of interaction between the northern and southern structures. Moreover, even in the absence of A1758S, the resulting simulations proved to be sufficient to explain the internal dynamics of A1758N itself.

A1758S is less studied than its more massive northern companion. Nevertheless, A1758S is suspected to be also a binary merger in its own turn. From its X-ray morphology, \cite{David2004} proposed that the two substructures within A1758S might be undergoing a nearly head-on collision. Based on only 27 spectroscopic members, \cite{Monteiro2017} proposed a scenario in which the two substructures within A1758S would be colliding along a direction which is closer to the line of sight than to the plane of the sky. This collision would probably be in an early stage of the merger. Additionally, there is an X-ray group found by \cite{Haines2018} that was spectroscopically confirmed to be associated with A1758N, and adds a further substructure to this already complex system. 

{In fact, a line-of-sight velocity offset of $-1200$\,km\,s$^{-1}$ between A1758S and A1758N was measured in \cite{Monteiro2017}, with A1758S at $z=0.274$ and A1758N at $z=0.279$. More recently, \cite{Bianconi2020} combined redshifts obtained by \cite{Monteiro2017} with their own redshifts from the Arizona Cluster Redshift Survey (ACReS), and additional redshifts from \cite{Haines2013}, to conduct a dynamical analysis of A1758. They identified members of A1758S and of A1758N (within $r_{200}$), with central redshifts of 0.27386 for A1758S and 0.27879 for A1758N. Consequently, the calculated line-of-sight velocity offset for A1758S relative to A1758N would be $-1156$\,km\,s$^{-1}$. \cite{Bianconi2020}, showing the redshift as a function of declination for spectroscopic members of A1758, concludes that the member galaxies of A1758S can be clearly distinguished from A1758N.}

Regardless of the internal structure of A1758S, the possible interaction between A1758N and A1758S remained unclear at first. From X-ray observations with \textit{Chandra} and \textit{XMM-Newton}, \cite{David2004} had put forth the possibility of the N and S clusters being gravitationally bound, although at that time there were yet no direct indications of interaction between them.

\cite{Botteon2018} combined LOFAR (LOw Frequency ARray) observations with archival data from GMRT (Giant Metrewave Radio Telescope) and VLA (Very Large Array). They suggested a possible low surface brightness radio bridge connecting A1758N and A1758S. Furthermore, analysis of X-ray \textit{Chandra} data points to the unrelaxed state of the outskirts of the clusters and, possibly, to some compression and heating of the region between the pair. From this putative equatorial shock, \cite{Botteon2018} estimate a Mach number of $\mathcal{M}_{kT} = 3.0^{+1.4}_{-1.0}$ from the temperature jump.

Subsequently, \cite{Schellenberger2019} analysed shocks in the edges of A1758N and of A1758S. Interestingly, they found a temperature structure around the NW cluster similar to the missing bow shock predicted by the simulation in \cite{Machado2015}. However, no conclusive evidence was found for shocked gas between A1758N and A1758S.

More recently, \cite{Botteon2020} confirmed the detection of the giant radio bridge that had been tentatively suggested earlier \citep{Botteon2018}. The bridge is clearly seen in the LOFAR image, and the radio and X-ray emissions are correlated.

Once the existence of a subtle but detectable bridge has been established, the presence of A1758S can no longer be overlooked in a consistent model of A1758. Specifically, the question arises about what dynamical configuration could have brought A1758S to its current position. Thus, dedicated simulations are needed in order to include the arrival of A1758S. In this paper, we explore a suite of different scenarios which could have led to the observed configuration of the A1758 system. In particular, we wish to evaluate whether those possible scenarios would be consistent with a tenuous bridge between the clusters. In other words: is it conceivable, from simulations, that A1758S could have arrived at its present location without inducing very substantial X-ray excess due to compression?

The diffuse radio emission detected between the clusters \citep{Botteon2018, Botteon2020} is understood to be due to synchrotron radiation from relativistic electrons in the presence of amplified magnetic fields. However, in this paper, we will not simulate radio emission. Furthermore, in the context of such a radio bridge, it is expected that some additional X-ray emission might also arise from inverse Compton scattering \citep{Brunetti2020}. This mechanism will not be the focus of the current paper either. Instead, we will focus on the thermal emission only. We will explore possible enhancements in X-ray emission, but only due to gas compression, not due to additional non-thermal mechanisms.

This paper is structured as follows. In Section~2, we describe the initial conditions of the simulations and also the proposed scenarios. In Section~3, we present the results for the radial approach and for the other scenarios, as well as the analysis of the shocks and the X-ray mocks. Summary and conclusions are given in Section~4. At a redshift of $z = 0.278$, 1\,arcsec corresponds to 4.25\,kpc for a cosmology with $\Omega_{\rm m}=0.27$, $\Omega_{\Lambda}=0.73$ and $H_0 = 70\,{\rm km}\,{\rm s}^{-1}\,{\rm Mpc}^{-1}$.

\vfill

\section{Simulation setup}

The simulations presented in this paper build upon the previous results in \citet[][hereafter M15]{Machado2015} and in \citet[][hereafter MO17]{Monteiro2017}. Our simulation of A1758N is thoroughly explained in M15. With the weak lensing masses of MO17, we had performed an updated simulation -- this is the version taken to be the fiducial one, which is the basis for the composite model developed in the present paper. However, the two versions produced qualitatively similar outcomes. In this paper, we will focus on the arrival of A1758S. The MO17 model of A1758N will be taken as is. Thus, the same A1758N initial conditions will be applied, and A1758S will be added in the ways described in Section~\ref{sec:scenarios}

\subsection{Initial conditions}

As a first approximation, we will ignore the possible merging taking place within A1758S: it will be modelled as a single structure, because our focus for the moment is on its probable orbit, not on its internal structure. The total mass of A1758S determined in MO17 is $4.96^{+1.08}_{-1.19} \times 10^{14}\,{\rm M}_{\odot}$.

To create the initial conditions for an idealised cluster representing A1758S, we followed the same procedures outlined in M15. Here we give a brief summary of the main properties. The cluster is represented by an initially spherical distribution of DM and gas. The virial mass is set to $5.2 \times 10^{14}\,{\rm M}_{\odot}$, and the adopted gas fraction is 15 per cent. The DM halo follows a \cite{Hernquist1990} profile with a scale length of 400\,kpc. The gas follows a \cite{Dehnen1993} profile with the parameter $\gamma=0$, which implies a core that resembles a $\beta$ model. The choice of a gas scale length of 210\,kpc, together with the halo potential, leads to a central temperature of $\sim$6.4\,keV, due to the imposition of hydrostatic equilibrium. Further details of the numerical realisation can be found in \cite{Machado2013}. The initial conditions for A1758S are generated with a resolution of $10^6$ particles for the DM and $10^6$ particles for the gas, using the clustep\footnote{\url{https://github.com/ruggiero/clustep}} code \citep{Ruggiero2017}. In keeping with the procedures in M15, the simulations are carried out with the Gadget-2 code \citep{Springel2005} with a gravitational softening length of 5\,kpc.

This initial condition created to represent A1758S is then joined with the initial condition of the A1758N merger to produce the complete $t=0$ snapshot of the actual simulation. The separations and relative velocities will be specified in Section~\ref{sec:scenarios}.

For completeness, it should be mentioned that the A1758N initial conditions are composed of 2 clusters (NW and NE), each created with analogous procedures and with parameters given in MO17. Likewise, both NW and NE are each composed of $10^6$ particles for the DM and $10^6$ particles for the gas.

\begin{figure}
\centering
\includegraphics[width=\columnwidth]{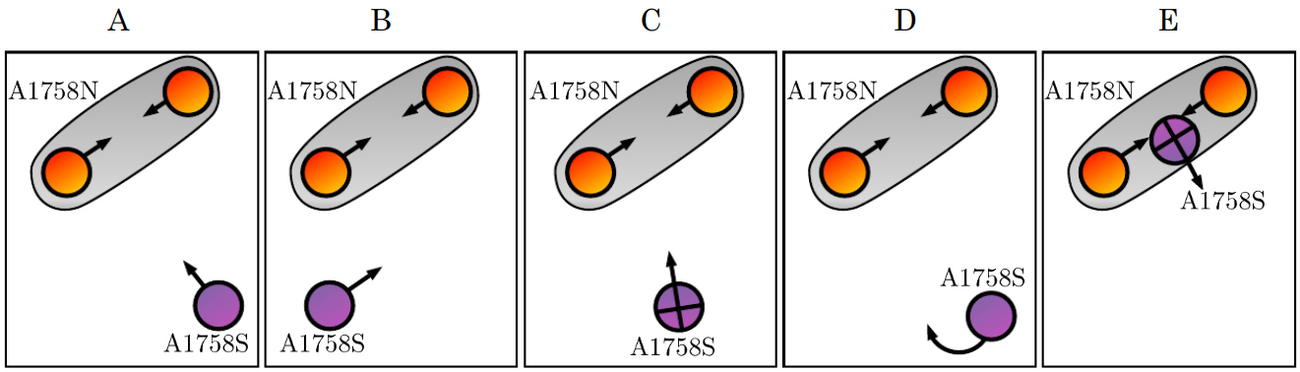}
\caption{Schematic representation of the initial conditions in different scenarios: (A) radial approach, (B) tangential approach, (C) vertical approach, (D) post-apocentric passage, and (E) outgoing.}
\label{fig01}
\end{figure}

\subsection{Scenarios}
\label{sec:scenarios}

Within the A1758N collision, the NW and NE clusters pass by one another at a pericentric separation of only 80\,kpc. Approximately 0.3\,Gyr later, the so-called ``best instant'' of the simulation is reached. This is the moment when the gas morphology as well as other quantitative constraints are satisfied. The challenge of setting up a three-body interaction is to have A1758S arrive at the desired position precisely at that moment. In all the proposed scenarios, we assume that A1758S is undergoing its first passage.

\begin{table*}
\caption{Parameters of the initial conditions. Position and velocity coordinates of A1758S are given with respect to the centre of mass of A1758N.}
\centering
\begin{tabular}{ccccccccc}
\hline
Model & $x$ & $y$ & $z$ & $|\mathbf{r}|$ & $v_x$ & $v_y$ &  $v_z$ & $|\mathbf{v}|$ \\
 & (kpc) & (kpc) & (kpc) & (kpc) & (km\,s$^{-1}$) & (km\,s$^{-1}$) & (km\,s$^{-1}$) & (km\,s$^{-1}$) \\
\hline
Radial approach (A) &   959   & $-$2966 &    0 & 3117 &     0   &      0  &       0 &    0 \\
Tangential approach (B) & $-$2600 & $-$3100 &    0 & 4043 &  2000   &      0  &       0 & 2000 \\
Vertical approach (C) &   859   & $-$2657 & 3350 & 4361 &     0   &      0  & $-$2000 & 2000 \\
Post-apocentric (D) &  3000   & $-$1964 &    0 & 3586 & $-$1200 &  $-$500 &       0 & 1300 \\
Outgoing (E) &     0   &     0   & 2000 & 2000 &   640   & $-$2000 &  $-$450 & 2148 \\
\hline
\end{tabular}
\label{tab:ic}
\end{table*}

In order to clarify the spatial coordinates of the initial conditions, we need to recapitulate the A1758N setup. In that collision, the NW and NE clusters were set initially 3\,Mpc apart along the $x$-axis, but also shifted by an impact parameter of $b=350$\,kpc along the $y$-axis. The initial relative velocity was $-1500$\,km\,s$^{-1}$ parallel to the $x$-axis direction. However, it should be noted that (unless otherwise stated) the figures in this paper display the result of the simulation rotated by an angle around the $z$-axis. This is done merely to have the simulation result match the position angle of the observations. It is not an inclination with respect to the line of sight, and thus the $xy$ simulation plane remains parallel to the plane of the sky.

Having set the stage of the A1758N collision, we may now bring A1758S into the picture. The initial position of A1758S will be given with respect to the centre of mass of A1758N. Likewise, the initial velocity of A1758S will be given as a velocity relative to A1758N.

In order to explore a wide variety of possible orbits, we have set up five distinct scenarios describing how A1758S could have arrived at its present position. These configurations are meant to be representative of plausible regimes for the initial conditions.

In the observed system, A1758S lies at $\sim$2.2\,Mpc roughly to the south of A1758N (actually about $15^{\circ}$ to the west of the north-south direction). Thus, in each of the five simulations, the A1758S cluster must be required to arrive at that desired location specifically at $t=1.47$\,Gyr. Here, it is not an issue of redefining the time coordinate arbitrarily, because that specific moment (by any label one would call it) is the intrinsic snapshot when the appropriate A1758N morphology is attained. Earlier or later snapshots would not be suitable. The final three-body choreographies are achieved by trial and error. Since the clusters are considerably far apart, one may obtain an initial estimate of a plausible orbit via a two-body calculation by replacing both and A1758N and A1758S by point masses. This provides a rough educated guess, which then needs to be fine-tuned, first with low-resolution DM-only simulations, which are inexpensive, and finally with the full high-resolution clusters. This was done separately for each of the five scenarios. The coordinates of initial positions and initial velocities are given in Table~\ref{tab:ic}. A sketch of each scenario is drawn in Fig.~\ref{fig01}. The initial conditions of each scenario are individually explained in the following:

\paragraph{(A) Radial approach} This is the simplest assumption. A1758S falls radially towards A1758N. In the initial condition, A1758S is placed at rest, at an initial separation of 3117\,kpc. Even though in the default model A the initial velocity is effectively zero, the arrow sketched in Fig.~\ref{fig01} is meant to represent what happens as soon as the simulation starts. In Section~\ref{sec:radial}, variants of model A with higher (non-zero) initial velocity are also explored. 

\paragraph{(B) Tangential approach} In this scenario, the initial velocity of A1758S is parallel to the merger axis of A1758N, rather than being perpendicular to it. In the initial condition, A1758S is placed at an initial separation of 4043\,kpc, with an initial velocity of 2000\,km\,s$^{-1}$ entirely in the tangential direction.

\paragraph{(C) Vertical approach} In this case, A1758S starts with a separation of 2792\,kpc on the $xy$ plane, but it is also raised 3350\,kpc above the $xy$ plane. The initial velocity is 2000\,km\,s$^{-1}$ pointing entirely down in the vertical direction. In Fig.~\ref{fig01}, the cross represents the vertical velocity going into the page. The sketched arrow should be understood as a velocity component that will develop towards A1758N once the simulation starts.

\paragraph{(D) Post-apocentric} In this scenario, A1758S starts at a separation of 3586\,kpc, moving initially away from A1758N in a general direction intermediate between radial and tangential. The small velocity of 1300\,km\,s$^{-1}$ is such that it soon reaches its apocenter and then turns into an incoming trajectory. With some graphic liberty, the curved arrow in Fig.~\ref{fig01} should be seen not as the $t=0$ velocity vector, but rather as the qualitative shape of the trajectory around the point of return.

\paragraph{(E) Outgoing} In this scenario, A1758S starts moving outwards, radially away from A1758N. A1758S could neither be placed between NW and NE, nor be made to cross that region, because this would disrupt too dramatically the already settled NW--NE interaction. The solution was to raise A1758S 2000\,kpc above the centre of mass of A1758N, and set it with an outwards velocity of 2148\,km\,s$^{-1}$, represented by the arrow sketched in Fig.~\ref{fig01}. The into-the-page cross depicts the downward vertical velocity component that will develop as soon as the simulation starts.

\section{Results}

\subsection{Radial approach}
\label{sec:radial}

In this section, we present the results of the radial approach, which is the simplest assumption. In the default version of scenario~A, the A1758S is released with zero velocity and then falls toward A1758N.

\begin{figure*}
\centering
\includegraphics[width=\textwidth]{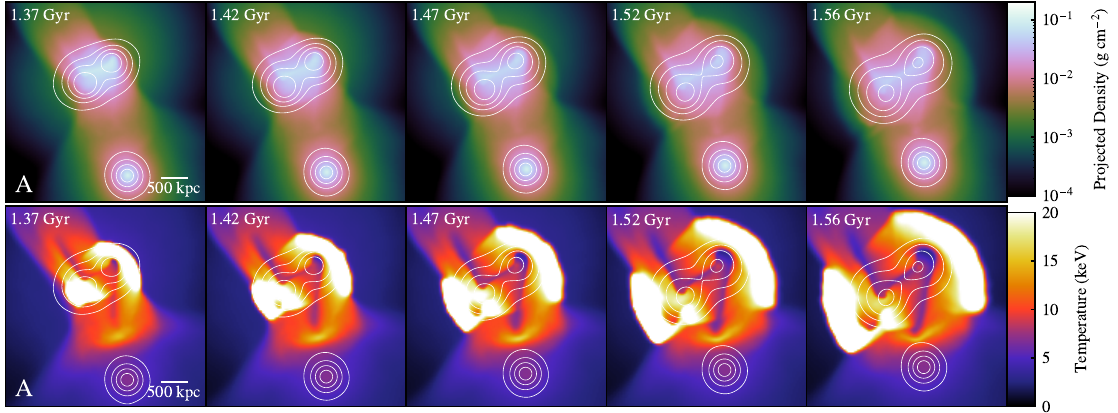}
\caption{Time evolution of model A around the best time ($t=1.47$~Gyr): projected gas density (top row); emission-weighted temperature (bottom row). The white contours represent total projected mass.}
\label{fig02}
\end{figure*}

Fig.~\ref{fig02} shows the time evolution of scenario~A around the best-matching instant, which is the middle frame ($t = 1.47$\,Gyr). The frames are 3.5\,Mpc wide and display the projected gas density and the emission-weighted temperature in the plane of the orbit, with the position angle having been rotated to match the orientation of the observations.

Regarding the internal morphology of A1758N, the best moment had been selected based on the criteria presented in M15, mainly the separation between the DM peaks of the NW and NE structures, and the DM/gas dissociation. Now, the best moment corresponds, additionally, to the time at which A1758S arrives at a position 2.2\,kpc away from A1758N. The initial conditions were retroactively constructed in order to ensure this synchronicity.

\begin{figure}
\centering
\includegraphics[width=0.5\columnwidth]{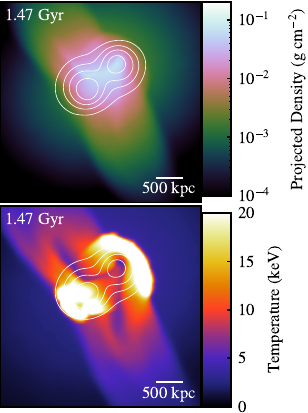}
\caption{Density and temperature maps of the A1758N-only simulation, i.e.~in the absence of A1758S. The white contours represent total projected mass.}
\label{N-only}
\end{figure}

In spite of the gradual approach of A1758S, the time evolution in Fig.~\ref{fig02} reveals that the morphology of A1758N remains sufficiently undisturbed with respect to the original model (see Fig.~\ref{N-only}). Mild increase in gas densities can be seen in the region between A1758S and A1758N.

For comparison, we display the A1758N-only simulation model in Fig.~\ref{N-only}. This corresponds to the best model of A1758N in MO17, i.e.~in the absence of A1758S. This model serves as the reference for further comparisons to evaluate the impact of including A1758S.

The impact of the arrival of A1758S seems more noticeable in the temperature maps of Fig.~\ref{fig02}. In the intermediate region between the clusters, gas is heated to around 10\,keV. This temperature structure in the intermediate region was not seen in the A1758N-only simulation (Fig.~\ref{N-only}). The very hot outwards-propagating shock fronts due to the internal collision within A1758N were already present in the earlier model and remain quite similar.

\begin{figure}
\centering
\includegraphics[width=\columnwidth]{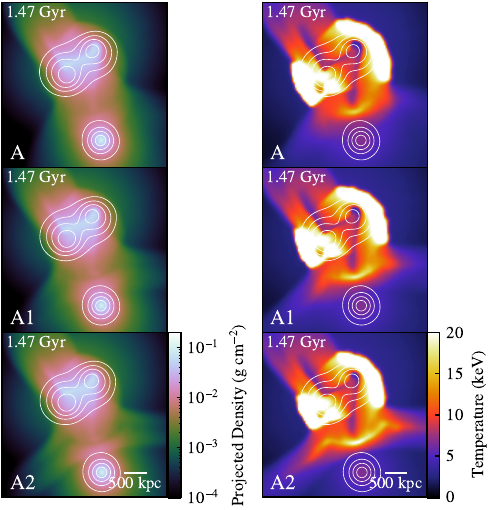}
\caption{Model A compared to the higher-velocity variants A1 and A2, at the best time: projected gas density (left column); emission-weighted temperature (right column). The white contours represent total projected mass.}
\label{fig03}
\end{figure}

To explore the possible effects of different velocities, we create two variants of the default scenario A. These variants are model A1, with an initial velocity of 632\,km\,s$^{-1}$, and model A2, with an initial velocity of 1391\,km\,s$^{-1}$. These are the initial relative velocities of A1758S, and they point towards the centre of mass of A1758N. Because of the higher velocities, the initial condition of A1758S naturally needs to be placed farther away. The initial separations are thus 3836\,kpc for model A1, and 4795\,kpc for model A2.

\begin{figure*}
\centering
\includegraphics[width=\textwidth]{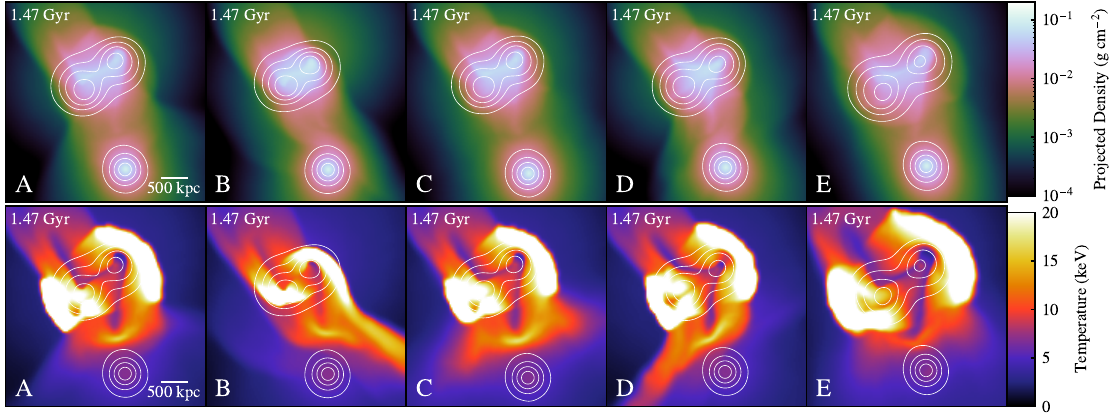}
\caption{Comparison of all models (A to E) at the best time: projected gas density (top row); emission-weighted temperature (bottom row). The white contours represent total projected mass.}
\label{fig04}
\end{figure*}

A comparison between the models A, A1 and A2 is shown in Fig.~\ref{fig03}. Projected gas density and emission-weighted temperature are shown only at the best time for the three models. Looking at the intermediate region between the A1758N and A1758S clusters, it is clear that the compression of the gas increases in the higher velocity models. The density enhancement that was subtle in model A becomes slightly more noticeable in A1 and much more so in A2. Likewise, the temperature enhancement also becomes more pronounced in models A1 and A2, with the temperature of the shocked gas increasing progressively from $\sim$10 to $\sim$12 to $\sim$15\,keV in models A, A1 and A2, respectively. Nevertheless, the original shock fronts around A1758N itself remain quite similar to the default model A.

It should be noted that the velocities quoted earlier refer to the initial conditions. In model A, the A1758S cluster was accelerated from rest to 1718\,km\,s$^{-1}$. This is the final relative velocity (i.e., at the best moment). Similarly, the final relative velocities of models A1 and A2 are, respectively, 1714\,km\,s$^{-1}$ and 2668\,km\,s$^{-1}$. In the case of A1, the A1758S cluster had to cover a longer distance; but, starting with a non-zero velocity, it spent most of its trajectory with higher velocities than in model A. This led to cumulatively more gas compression, although at the specific instant $t = 1.47$\,Gyr the instantaneous velocities in A and A1 happen to be similar. For A2, on the other hand, the final velocity is substantially higher.

\vfill

\subsection{Other scenarios}

Having established the properties of the more straightforward radial scenario A, in this section we will compare it to the remaining scenarios B--E.

Fig.~\ref{fig04} shows the results for models A--E, displaying the projected gas density and the emission-weighted temperature. Only the best moment is shown. In all cases, A1758S has approximately the same position, $\sim$2.2\,Mpc from A1758N -- as can be seen both in the density and in the temperature maps because its position coincides with the peak of projected mass in the south, i.e.~the center of the countour lines in Fig.~\ref{fig04}. The central temperature of A1758S is $\sim$6\,keV.

Regarding the internal structure of A1758N, one can discern some differences in the panels of Fig.~\ref{fig04}. Although the global morphological features are roughly preserved in all cases, the density maps do exhibit some variation depending on the model. This can be understood as a natural consequence of the additional gravitational influence of A1758S. The approach of A1758S will inevitably perturb the original NW-NE orbit to some degree.

Likewise, the overall temperature structure of A1758N is perturbed as well -- more evidently so than the density maps. The shock fronts in such a configuration are understood to propagate with a higher velocity than the velocity of the cluster themselves \citep[][M15]{Springel2007}. For this reason, the shape and extent of those A1758N shock fronts are somewhat more time-sensitive than other mechanical features. Thus, a relatively small gravitational perturbation might alter the exact orbit of the NW-NE pair in such a way as to slightly delay or anticipate the initial propagation of the shock front. This perturbation would shift the gas density and DM peaks by a small extent, but would change the global shape of the shock fronts more noticeably.

In any case, in this paper we do not aim to refine the internal details of the A1758N model to accommodate the new perturbation. Rather, we assume that the A1758N remains qualitatively similar and focus on the intermediate region between the clusters. In the following, a description of each scenario B--E is presented individually.

In scenario B, the A1758S cluster moves tangentially at first, and then is slightly bent due to the gravitational attraction of A1758N. By the time A1758S reaches its desired position, its velocity is 2191\,km\,s$^{-1}$. The resulting gas compression is perpendicular to the motion of A1758S and is seen as something akin to proper bridge, appearing to extend from one cluster to the other (density B panel of  Fig.~\ref{fig04}). Consequently, the heated gas in the temperature is also oriented in the same perpendicular direction. This shocked gas modifies (and partially merges with) the southwestern edges of the NW shock front. The shape of the model B shock is qualitatively dissimilar from that of model A, as they were indeed meant to be as orthogonal to one another as possible, thus exploring the extremes of two possible regimes. Nevertheless, the values of the temperatures in the shocked regions of A and B are roughly in the same range. Since A1758S comes from the east in this model, it spends most of its approach in the vicinity of NE, or at least on its general side. Thus, the NE shock front is the most perturbed and altered with respect to the default model.

In scenario C, the A1758S cluster starts above the plane, coming down. Its initial velocity vector is purely vertical, but the gravitational attraction of A1758N naturally adds a velocity component towards the northern structure. In a sense, the resulting model C is a three-dimensional version of the orbit of model A, in which the A1758S is also coming down while it moves towards A1758N. However, the final motion is still predominantly along the line of sight. The final velocity is 2606\,km\,s$^{-1}$, but the $v_z$ component is 2368\,km\,s$^{-1}$. The projection of the resulting  shock is less straightforward to envision in such a case. (In fact, projections of complicated three-dimensional temperature structures tend to be quite non-intuitive in simulated cluster mergers;  \citealt{Machado2022}). The C temperature panel of Fig.~\ref{fig04} indicates the general morphology of the intermediate shock is closer to A than to B, as would be expected. The velocities are such that the gas was heated to $\sim$15\,keV. In the C density panel of Fig.~\ref{fig04}, the supposed compressed gas is practically not noticeable. This might be understood if the $xy$ plane is though of as a projection of an inclined shock whose major contribution is parallel, nor perpendicular to it. In this configuration, an overdensity ahead of the incoming cluster would be diluted in projection.

In scenario D, the A1758S cluster starts at a moment shortly prior to the apocentric passage, and then turns for an incoming orbit with generally oriented velocity. This scenario can be considered as a general intermediate version between A (radial) and B (tangential) -- but coming from the other side, i.e.~from the western side. This is the reason why the D temperature panel of Fig.~\ref{fig04} vaguely resembles a flipped version of B. In this case, the NE was not affected as in B; and the NW proved more resilient, since it was more to the north, i.e.~farther away from the perturber than NE was in model B. The resulting shock of scenario D also has maximum temperatures of about $\sim$15\,keV. The bridge in the D density panel of Fig.~\ref{fig04} is more complex, as would be expected from an orbit with more varied velocity components, including a point of return in the orbit. Still, this resulting density bridge can also be regarded qualitatively as a flipped version of model B. The final velocity of model D is 1864\,km\,s$^{-1}$.

In scenario E, the A1758S cluster starts directly above the centre of mass of the A1758N cluster, with an outwards velocity. As it moves radially outward, it also gains a downward velocity component, as it is pulled by the gravitational attraction of A1758N. It is interesting to see that the NW and NE shock fronts look more developed in the E temperature panel of Fig.~\ref{fig04} than in any of the other scenarios. As the initial condition of A1758S was placed instantaneously above A1758N, initially A1758S attracts both NW and NE towards itself, thus bringing them together slightly earlier, and precipitating the NW-NE collision to occur sooner by a short amount of time. The consequence in the density map is not great. But the NW and NE shock fronts look like a later version of the original model. Focusing on the intermediate region between the northern and southern clusters, the E density panel of Fig.~\ref{fig04} exhibits only a very tenuous irregular filament which is seen in other models, but no compression per se. This is understandable, since the A1758S cluster is moving outwards. In the E temperature panel of Fig.~\ref{fig04}, the gas below $\sim$12\,keV would seem to be more related to disturbances in the outer edges of the northern fronts, than to the motion of A1758S itself. The velocity of A1758S in this scenario is 1895\,km\,s$^{-1}$. The predominant component is vertical, with $v_z = -1779$\,km\,s$^{-1}$. In the $xy$ plane, the velocity component of 655\,km\,s$^{-1}$ points away from A1758N. In this sense, model E is the exception among the five studied scenarios: it is the only outgoing one.

A direct side-by-side comparison between simulated and observed temperature maps would be challenging. Instead, we give here a brief description of the approximate temperature values encountered in observations. In \cite{Schellenberger2019} temperatures in excess of 13\,keV are seen in the region slightly to the south of A1758N (their figure~3). This is generally consistent with our models A--D, which display temperatures peaking at roughly 15\,keV in the intermediate region between A1758N and A1758S. The possible exception to this agreement would be model E, where the temperature is slightly lower (Fig.~\ref{fig04}). In \cite{Botteon2018}, temperatures in excess of 13\,keV are also seen in the intermediate region (at least in the low signal-to-noise temperature map of their figure~A2). This is again generally consistent with the temperature maps of our models A--D (Fig.~\ref{fig04}), but a more quantitative comparison will be made in the next subsection.

\vfill

\subsection{Shocks}

\begin{figure}
\centering
\includegraphics[width=\columnwidth]{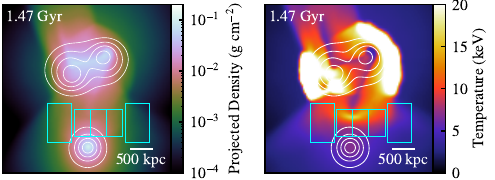}
\caption{Illustrative example of the regions used to measure the temperature between the clusters, following \cite{Botteon2018}. For convenience, this figure is drawn with a different position angle with respect to the other figures in the paper.}
\label{fig05}
\end{figure}

In this section, we investigate how the temperature structure might help interpret the meaning of the possible shocks.

First, we quantify the temperature profile in a manner that is inspired by the so-called transversal profile measured by \cite{Botteon2018}; their figure~9. In order to do so, we define five rectangular regions, as illustrated in the example of Fig.~\ref{fig05}. For convenience when drawing the rectangles, the snapshots were rotated around the $z$ axis. Regions 1 and 5 have dimensions 574\,kpc $\times$ 910\,kpc; regions 2 to 4 have dimensions 385\,kpc $\times$ 652\,kpc. Within each region, we take the mean and the standard deviation of the emission-weighted projected temperature. This is done for each of the five scenarios. In each scenario, the set of regions is actually shifted sideways by a given arbitrary amount along the horizontal axis of Fig.~\ref{fig05}. This is done in a way to have the highest temperatures land within regions 2 or 3, as is the case in the observations of \cite{Botteon2018}. This freedom is justified, since the morphology of the temperature structure varies from model to model, and thus it would be difficult to demand an absolute positioning of the presumed shocked gas. This measurement can be understood both as a rough attempt at observational comparison, but also as a theoretical characterisation of the profiles of each scenario.

\begin{figure}
\centering
\includegraphics[width=\columnwidth]{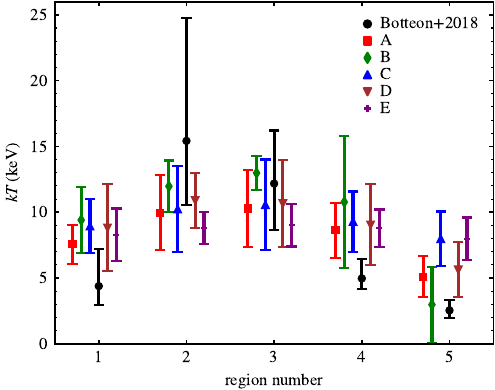}
\caption{Temperatures measured within regions such as in Fig.~\ref{fig05}. The black points are taken from \cite{Botteon2018}. Within each region, the points are slightly shifted sideways, for clarity.}
\label{fig06}
\end{figure}

The resulting transversal temperature profiles are shown in Fig.~\ref{fig06}. The black circles in Fig.~\ref{fig06} are temperature values taken from \cite{Botteon2018}. The other symbols correspond to our scenarios A--E. Within each region, the points were shifted for clarity. The global result of these measurements is that, even with a wide variety of morphologies, the simulated temperatures fall in the same approximate ranges as the observational results -- in some cases, with a borderline agreement of more than 1$\sigma$. However, the peak temperature (i.e.~the temperature increase from region 1 to 2, followed by a decline) is more pronounced in the observation than in the simulations. In particular, scenario E shows no suggestion of a peak. In view of these simulated profiles, it seems difficult to argue for a meaningful equatorial shock \citep{Ha2018} in our scenarios. Thus, we refrain from estimating Mach numbers under such assumption.

Instead, if we understand the heated gas in the intermediate region as shocks moving ahead of the incoming A1758S, we may estimate their Mach numbers in the usual ways. The assumption here is that the direction of the propagation of the shock front is perpendicular to the surface of temperature discontinuity (or rather, its projection onto the $xy$ plane). For example, in scenarios A and C, the direction of propagation of the shock front would be radially towards A1758N. In scenarios B and D, it would be perpendicular to the predominant direction of the heated gas.

\begin{table}
\centering
\caption{Mach numbers estimated via the temperature jump ($\mathcal{M}_{kT}$) and via the shock velocity ($\mathcal{M}_v$).}
\begin{tabular}{ccc}
\hline
model & $\mathcal{M}_{kT}$ & $\mathcal{M}_v$ \\
\hline
A  & 1.8--1.9 & 1.2 \\
A1 & 2.0--2.1 & 1.2 \\
A2 & 2.1--2.2 & 1.5 \\
B  & 1.6--1.7 & 1.4 \\
C  & 1.7--1.9 & 1.1 \\
D  & 1.7--1.8 & 1.9 \\
\hline
\end{tabular}
\label{tab:mach}
\end{table}

One method to estimate the Mach number is using the amplitude of the temperature discontinuity:
\begin{equation}
\frac{T_{2}}{T_{1}} = \frac{5\mathcal{M}_{kT}^{4}+14\mathcal{M}_{kT}^{2}-3}{16\mathcal{M}_{kT}^{2}},
\end{equation}
where $T_1$ is the pre-shock temperature and $T_2$ is the post-shock temperature \citep{Landau1959}. The Mach number estimated in this way is $\mathcal{M}_{kT}$; this is usually employed in observations but can also be applied to simulations. The other method consists in measuring the velocity of the shock front directly, from successive snapshots of the simulation. The location of the shock front is determined by the temperature discontinuity. Its advancement as a function of time gives the velocity $v_s$. However, the upstream gas may not be stationary \citep{Springel2007, Machado2013}. Rather, this gas ahead of the shock may have a velocity $u$ (in the same rest frame). Thus, the shock encounters the pre-shocked gas with an effective velocity of $v_s-u$. For this reason, the Mach number computed via the shock velocity is:
\begin{equation}
\mathcal{M}_v =  \frac{v_{s}-u}{c_{s}},
\end{equation}
where $c_{s}^{2}=\frac{\gamma k T}{\mu m_{p}}$ is the sound speed of the pre-shock gas. In simulations, the estimates of Mach number via the temperature jump ($\mathcal{M}_{kT}$) or via the shock velocity ($\mathcal{M}_v$) will generally coincide very well, if the merger configuration is that of a clean textbook shock, such as the `Bullet Cluster' \citep{Clowe2006} itself.
The Mach numbers estimated for our scenarios are presented in Table~\ref{tab:mach}, for both methods. They generally do not agree, with the temperature method almost always overestimating the Mach number, in comparison to the velocity method. An estimated confidence range is offered for the temperature method, due to the uncertainty in the choice of the region corresponding to the pre-shock gas. In fact, the temperature method is expected to be quite unreliable in this context, due to the complex temperature structure in the pre-shock region. Indeed, even though the so-called pre-shock has not yet been reached by the A1758S-induced shock, it is far from an undisturbed gas. On the contrary, it is a region quite perturbed by the A1758N-induced disturbances. The velocity methods would tend to be more reliable, but even considering the $u$ correction, it is still a complex configuration quite different from an idealised discontinuity. Scenario E is omitted from the Mach number analysis. Since that is an outgoing configuration, there would be no justification to interpret the hot gas in the intermediate region as a shock front. Indeed, visual inspection of the time evolution suggests that the hot region corresponds do the edges of the A1758N shocks dissipating as they expand. Overall, it is only possible to conclude that the Mach numbers of our scenarios may be in the approximate range 1--2, if the velocity method is to be relied upon.

\subsection{X-ray emission}

Finally, we resort to producing mocks of the X-ray emission in an attempt to make a more direct connection with observations.

In order to produce realistic X-ray mocks, we employ pyXSIM\footnote{\url{https://hea-www.cfa.harvard.edu/~jzuhone/pyxsim}} \citep{ZuHone2016}, which is a Python package for simulating X-ray emission from astrophysical sources, based on the PHOX algorithm \citep{Biffi2012, Biffi2013}. The use of pyXSIM is analogous to the procedures followed in \cite{Doubrawa2020} or \cite{Moura2021}. A fixed metallicity of 0.3\,Z$_{\odot}$ is assumed, as well as a hydrogen column density of $n_{\rm H} = 1.03 \times 10^{20}\,{\rm cm}^{-2}$ in the direction of A1758. The photons are projected along the line of sight, and the response from the detector is taken into account; in this case \textit{Chandra} ACIS-I, with an exposure time of 150\,ks \citep[as in][]{Botteon2018} and an energy band of 0.5--7.0\,keV.

\begin{figure}
\centering
\includegraphics[width=\columnwidth]{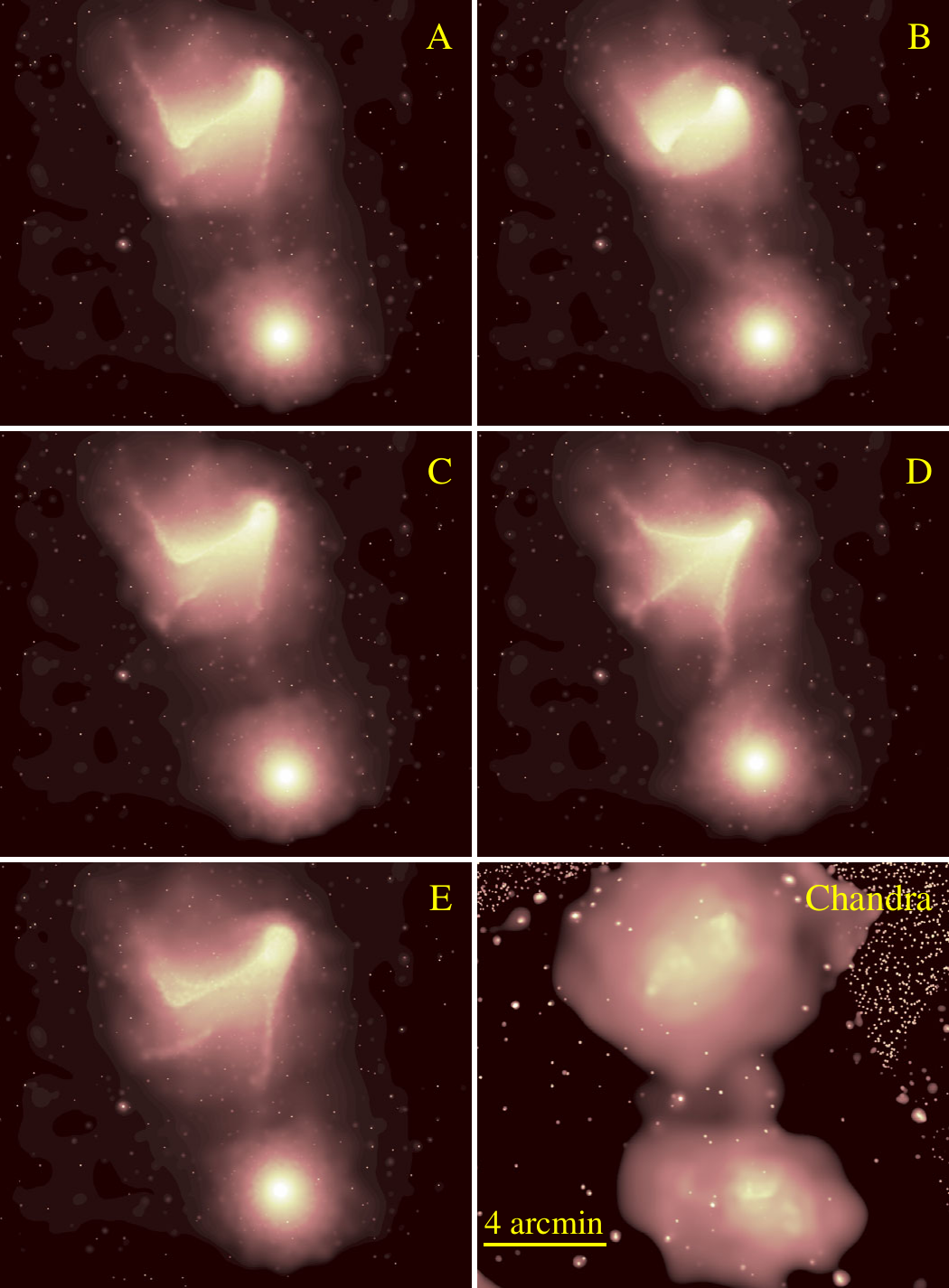}
\caption{X-ray mocks for all models (from A to E). The colours represent simulated X-ray counts smoothed with an adaptive kernel (see text), in logarithm scale. The bottom right panel shows the exposure-map corrected mosaic image of \textit{Chandra} exposures, also adaptively smoothed.}
\label{fig07}
\end{figure}

The simulated images are equivalent to the exposure-map corrected images from real \textit{Chadra} data, corrected from vignetting effects. The effect of CCD gaps of the simulated ACIS-I exposures were minimised by simulating a dithered observation. These mock images were further processed by using an adaptively smoothing, using a Gaussian kernel that increases in order to have a preset significance. We have used the \texttt{csmooth} tool from the \textsc{ciao} package from the \textit{Chandra X-ray Center} (CXC)\footnote{\url{https://cxc.cfa.harvard.edu/ciao/}}.

The resulting X-ray mocks are shown in Fig.~\ref{fig07} for models A--E. Contours of projected mass are not shown in Fig.~\ref{fig07}, to allow for a clearer inspection of subtle morphological features. Moreover, the location of the X-rays peaks with respect to the projected mass is qualitatively the same as in the case of gas density, which can be consulted in the previous figures. The main result of Fig.~\ref{fig07} is that a weak bridge in the intermediate region between A1758S and A1758N can be seen in the smoothed mock images, with somewhat different morphologies. The presence of these X-ray bridges shows that the mild over-densities seen in the gas density maps were indeed translated into the more realistic synthetic X-ray images. The fine-tuning of the internal sctructure of A1758N itself is not the main focus of the present paper. Yet, it is interesting to notice in Fig.~\ref{fig07} that model B exhibits perhaps the best morphological agreement with the \textit{Chandra} observation.

In order to compare our simulated images with the actual observations, we have downloaded the publicly available \textit{Chandra} observations, 5 ACIS-I (totalling 176.6\,ks) and one ACIS-S (59\,ks), with Obs Ids: 2213, 5772, 7710, 13997, 15538, and 15540. We have reprocessed all these exposures following the standard procedure using the \texttt{chandra\_repro} tool. Then, we have created a single broad band (0.5--7.0\,keV), exposure-map corrected image by merging all 6 exposures with the tool \texttt{merge\_obs}. Finally, as we did with the mock images, we have adaptively smoothed the \textit{Chandra} image, which is shown in the bottom right panel of Fig.~\ref{fig07}. The bridge between the North and South components is clearly seen.

\begin{figure}
\centering
\includegraphics[width=0.85\columnwidth]{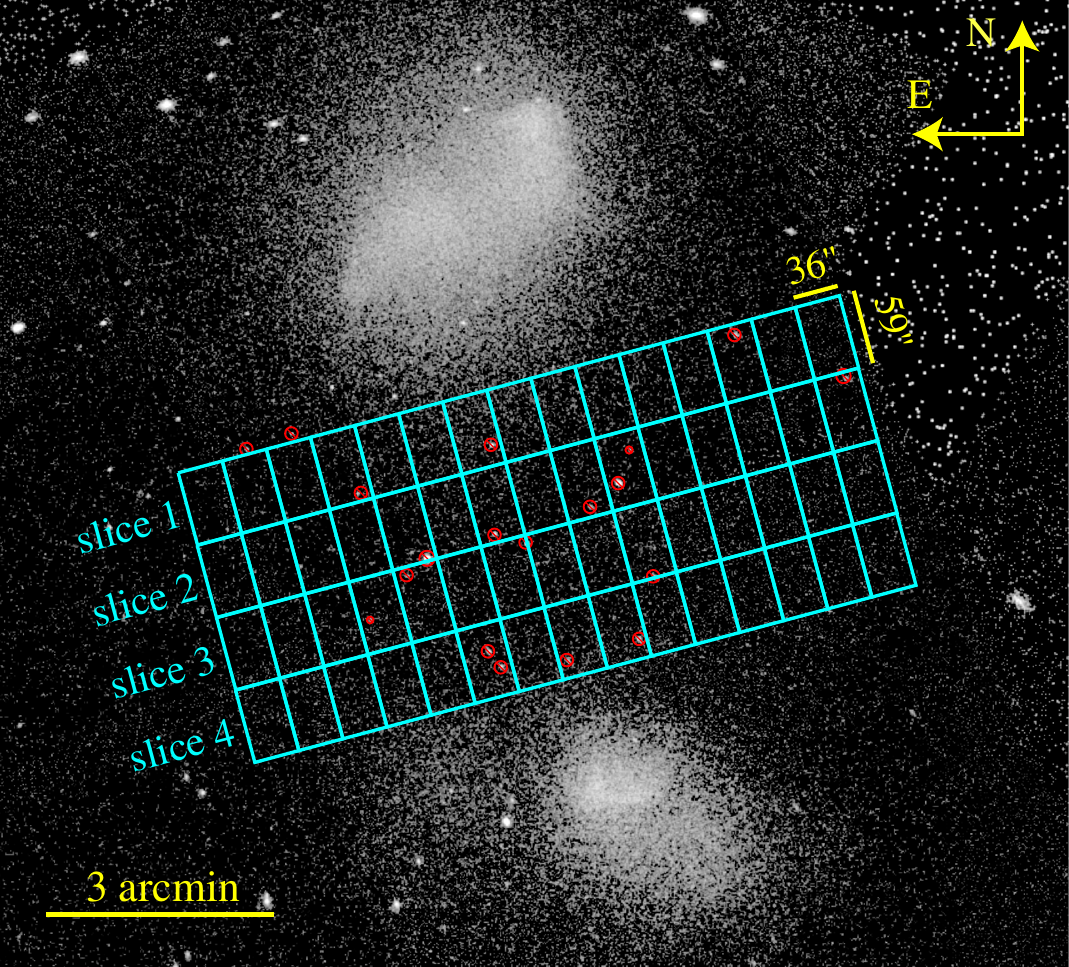}
\caption{Slices used to measure the X-ray surface brightness overlaid with the \textit{Chandra} exposure map corrected image. Each slice is composed of 15 rectangular regions. Each region is $36'' \times 59''$ (or 153\,kpc $\times$ 251\,kpc), rotated by $15^\circ$ counterclockwise. Red circles represent masked regions around point sources.}
\label{slices2}
\end{figure}

To further compare simulations with observation, we measured profiles of X-ray surface brightness across the region of the bridge. This is similar but not identical to the measurements in \cite{Botteon2020}. We defined 4 slices in a direction perpendicular to the line connecting A1758N and A1758S. These slices are shown in Fig.~\ref{slices2} with respect to the observed \textit{Chandra} image. Along each given slice, the X-ray surface brightness was measured within 15 successive rectangular regions, each of size 153\,kpc $\times$ 251\,kpc, with point sources masked. The ratio was computed between the excess in each region with respect to the average surface brightness of that entire slice. In this definition, if the bridge were absent, the profiles would be consistent with noise around the value of zero. This procedure was applied with the same techniques for the observed \textit{Chandra} image, as well as to the simulated models A--E.

\begin{figure}
\includegraphics[width=\columnwidth]{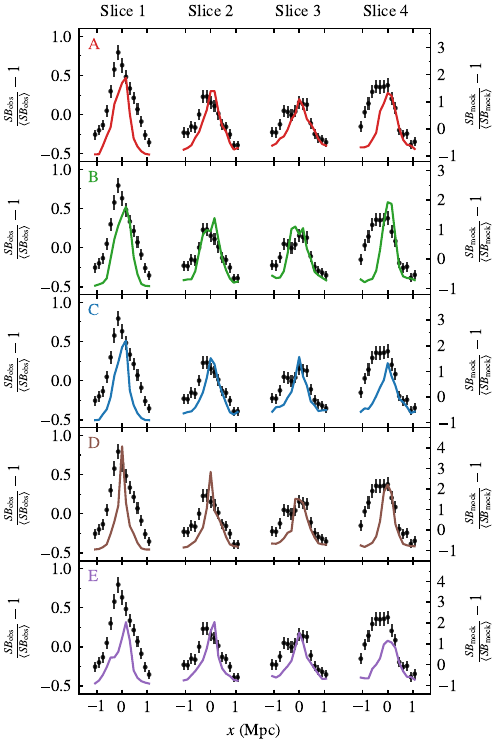}
\caption{Profiles of relative X-ray surface brightness measured along the slices defined in Fig.~\ref{slices2}. The black points correspond to measurements from the \textit{Chandra} image. The color lines represent the equivalent measurements using the mock images produces for the simulated models A--E. The `$x$' axis corresponds to the directions along the slices seen in Fig.~\ref{slices2}; in each case, $x=0$ means the central bin of that slice.  Note the different vertical scales in the righ-hand axes.}
\label{SB_profile_comp}
\end{figure}

\begin{figure}
\includegraphics[width=\columnwidth]{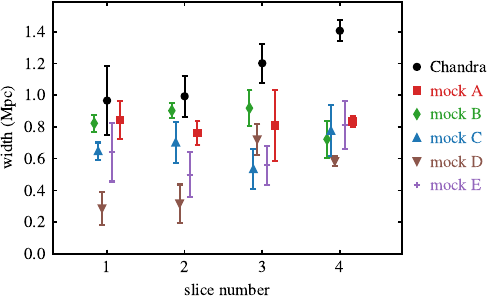}
\caption{Bridge widths obtained from the FWHM of the X-ray surface brightness profiles of Fig.~\ref{SB_profile_comp}. Within each slice, the points are artificially shifted sideways, for clarity.}
\label{widths}
\end{figure}

The resulting relative profiles of X-ray surface brightness are shown in Fig.~\ref{SB_profile_comp}, where the observational data (points) are compared to the mocks of each simulated model (lines). The peaks in the observational profiles are comparable to the ones measured in \cite{Botteon2020}. Peaks are detectable in all the 4 slices, but peaks 1 and 4 are somewhat higher, because those slices are closer to A1758N and A1758S, respectively. Thus the profiles of slices 1 and 4 are partially capturing contributions from the clusters themselves. On the other hand, the two intermediate peaks 2 and 3 are comparatively lower, since those slices cross the lower density regions between the two clusters. For this reason, slices 2 and 3 are the most meaningful when trying to characterize the bridge.

Regarding the mocks, the relative profiles of X-ray surface brightness are shown in each of the 5 panels of Fig.~\ref{SB_profile_comp}. The fact that peaks are clearly obtained in all cases indicates that the simulated X-ray bridge is in fact detectable (as was already seen morphologically in Fig.~\ref{fig07}). However, the amplitudes of the simulated peaks do not match precisely the amplitudes of the observed peaks. For a qualitative comparison to be made, the vertical ranges of Fig.~\ref{SB_profile_comp} were not kept fixed. This means that the excess X-ray emission is present in all simulations, but the absolute contrast seems to be overestimated in the simulations, by factors of typically $\sim$2--3 in models A, B and C; and as much as a factor of $\sim$4 in models D and E. These results might be sensitive to the arbitrary definition of the slices.

Nevertheless, we may consider the relative heights of the peaks in each slice of Fig.~\ref{SB_profile_comp}. We find that model B is the only one where the observed pattern is recovered: peaks 2 and 3 are slightly lower than both peaks 1 and 4. For models A and C, the first peak is indeed the highest of the four slices, but peak 4 is not meaningfully higher than the intermediate ones. Models D and E fail at this comparison.

Even if the absolute values of the peaks are not directly comparable, their widths give us some relevant estimation of how physically broad the bridges are. We obtained (both for observation and simulations) the full width at half maximum (FWHM) of each profile. Errors bars were estimated by considering the FWHM obtained allowing $\pm3\sigma$ from the data points. The resulting widths are shown in Fig.~\ref{widths}. One noticeable feature of this comparison is that all models fail at reproducing the width of slice 4. This discrepancy was already perceptible looking at the slices 4 in Fig.~\ref{SB_profile_comp}, where the observed profile is clearly wider than all the simulated ones. Slice 4 presumably encompasses relevant contributions from the X-ray emission in the vicinity of A1758S. The fact that all simulations underestimate the width of the X-ray contrast in that slice indicates the limitations of representing A1758S by one single structure.

Focusing on the remaining slices in Fig.~\ref{widths}, we find that model B seems to be the one that more closely approaches the \text{Chandra} values -- specially bearing in mind that regions 2 and 3 are the most meaningful to characterize the bridge region itself. Model A would be the next best match, since it is at least marginally consistent with three slices (1, 2 and 3). Model C is marginally consistent with two slices (1 and 2). Model D fails at all slices, while model E matches only slice 1 and none of the more relevant intermediate slices.

\section{Discussion}

A1758 is a complex galaxy cluster, composed of two major structures, A1758N and A1758S, each of which seems to be undergoing its own internal binary merger. The internal structure of A1758N had been modelled by dedicated hydrodynamical $N$-body simulations previously in M15 and MO17. The presence of A1758S had been ignored in those earlier works, because there was no strong evidence of interaction at the time. With the recent confirmation by \cite{Botteon2018, Botteon2020} of a radio bridge connecting A1758N and A1758S, it became necessary to consider the arrival of A1758S.

Theoretically, simulations generally predict strong compression in the early stages of a merger. Thus it becomes relevant to pose the question of whether A1758S could be falling towards A1758N without inducing very obvious enhancement in X-ray surface brightness between the two clusters. In this paper, we aimed to address this question.

We investigated five scenarios: radial approach (A), tangential approach (B), vertical approach (C), post apocentric (D) and outgoing (E). The motivation for choosing this set of orbits was to attempt to cover a wide variety of possible configurations. In a sense, models A, B and C somehow represent an orthogonal basis of initial velocities in the three linearly independent directions. Model D acts as a kind of generic or intermediate configuration, albeit confined to the $xy$ plane. Finally, model E is the exception, being the sole outgoing model. It should be noted that the directions of the final velocities are naturally different from the initial ones, in general. The difficulty in finding suitable models is the fine-tuning necessary to achieve the synchrony between the arrival of A1758S at its observed position, and the best moment of the already-established internal A1758N model. 

Within each proposed scenario, we offer in this paper only one instance of such an orbit, without sweeping the parameter space of possible velocities. If one particular model fails to meet the observational constraints, this does not mean that that entire family of models is automatically ruled out, because alternate versions with different velocities could in principle reconcile the simulation with the observation. Conversely, if one given model is shown to be satisfactory, this is sufficient to claim that that is a plausible scenario.

Additionally, in the present paper, we have not yet delved into the internal structure of A1758S itself. Rather, we focused on the interaction between A1758S as a whole and A1758N. Similarly we took the final MO17 model of A1758N to be the fiducial one. We commented on some of the minor morphological impacts that the arrival of A1758S induce on the NW and NE shock fronts, but without attempting to re-refine the A1758N model itself to accommodate for the new perturbation.

Another important caveat is that in the present exploration we have not attempted to satisfy the line-of-sight relative velocity between A1758N and A1758S, which is estimated to be of approximately $-1200$\,km\,s$^{-1}$ \citep{Monteiro2017, Bianconi2020}. Models C and E are the only ones which, by construction, have nonzero velocities along the line of sight. They end the simulation with $-2368$\,km\,s$^{-1}$ and $-1779$\,km\,s$^{-1}$, respectively.

The temperatures of the shocked gas between the clusters were found to be in the same approximate ranges as in the observations. This is an indication that the choices of initial velocities are not only plausible, but also specifically consistent with the known observational constraints. However, we did not interpret that temperature structure as an equatorial shock \citep{Ha2018}. Instead, we understood that the shock fronts in the incoming models are associated with the compressed gas ahead of A1758S. We estimated the Mach numbers directly from the advancing positions of the temperature discontinuity in successive snapshots, and obtained values in the range $\sim$1--2. This method uses the velocity of the shock front but also takes into account the correction due to the infalling pre-shock gas \citep{Springel2007}. The attempt to estimate the same Mach numbers using the method of temperature jumps revealed that this methods might be too unreliable in this case, due to the complex nature of the temperature structure in the surrounding gas. That gas has already been severely perturbed by the shock fronts emanating from the internal A1758N collision.

Shock-heating of the gas is a naturally expected phenomenon in pre-merging system, commonly predicted in simulations \citep[e.g.][]{Takizawa1999}, and also observed in many pre-merging clusters \citep[e.g.][among many others]{PaternoMahler2014, Akamatsu2016, Akamatsu2017}. If the projected separation between the two clusters is larger than the sum of their individual virial radii, temperature enhancements are generally not observed \citep[e.g.][]{AndradeSantos2015, Wegner2017}. In the case of A1758, it is conceivable that the intervening gas, having been previously perturbed by the internal A1758N collision, may be more prone to heating than it would otherwise be, if A1758N were a relaxed cluster. It is also interesting to note that if the edges of the expanding shock fronts of NW and NE were to be interpreted as an equatorial shock of their own, then they might contribute to boost the incoming shock due to the arrival of A1758S. In other words, an equatorial shock emanating from the A1758N merger axis would be parallel to the shock front of the incoming A1758S.

We produced X-ray mock images from the simulations. These mocks revealed that a bridge of X-ray emission is indeed predicted by the simulations. By measuring transversal profiles, we were able to quantify the excess in X-ray surface brightness in the region between A1758N and A1758S. The amplitude of this excess is somewhat overestimated in the simulations by factors of $\sim$2--3 in the best cases (models A, B and C); and by factors of as much as $\sim$4 in the other models. This means that the X-ray excess is seen more sharply in the simulations, when compared to the relatively attenuate profiles seen in the \textit{Chandra} data. A factor of $\sim$2--3 does not seem excessively discrepant, especially bearing in mind the highly idealised nature of these simulations, as well as the uncertainties in the processing of the observational data.

Nevertheless, in the X-ray surface brightness profiles, the relative heights of the peaks revealed additional results. Model B proved to be the only one which systematically follows the same pattern as the observational data, in having the two lowest peaks in the intermediate region.

Furthermore, the peaks in the X-ray surface brightness profiles were used to estimate the physical width of the bridge region. Model B was the one that most approached the observationally determined widths, followed by models A and C. Finally, regarding the X-ray mocks, it is interesting to notice that in the end model B produced the best-matching internal morphology of A1758N itself, even though that feature was not deliberately selected for.

An additional caveat is to highlight that magnetic fields and relativistic electrons were not taken into account in these simulations and we have not calculated radio emissions. The analyses of this paper were restricted to theoretical hydrodynamical quantities (temperatures and densities) and to synthetic X-ray observations. The simulations predict slightly overestimated enhancements in X-ray surface brightness. Only the thermal contribution was taken into account. Thus, our simulated X-ray emissions should be regarded as lower limits, since the possible inverse Compton contribution \citep{Brunetti2020} was not included. The possible bridges we attempted to investigate were bridges of gas density or bridges of (bremsstrahlung) X-ray emission. We did not investigate bridges of radio emission, nor X-ray due to inverse Compton emission.

Also, for simplicity, the A1758N model was kept on the plane of the sky, since this proved sufficient to attain a suitable model in M15. In other words, with the orbit of NW/NE kept on the plane of the sky, this gave rise to an adequate explanation of the internal A1758N merger. This does not mean that non-zero inclinations could be ruled out. In the five scenarios we chose to put forth in this paper, the initial velocities of A1758S were set, for simplicity, either fully on the $xy$ plane, or fully along $z$. But due to the gravitational attraction of A1758N, the models with initially vertical velocities ended up with generically three-dimensional velocities, although in the final configuration the predominant component was still along the line of sight (away from the observer in model C, and towards the observer in model E). In any case, it would be difficult to argue that the scenarios presented here are making quantitative predictions, for example, regarding specific values of velocity in the line of sight. Within each scenario, we offered one possible example in that regime, but without systematically sampling the parameter space.

\section{Conclusions}

The main conclusions of this paper may be summarised as follows:
\begin{itemize}

\item Densities: The proposed incoming scenarios are all generally consistent with mild gas overdensities in the intermediate region between A1758N and A1758S.

\item Temperatures: The compression due to the incoming A1758S gives rise to shock-heated gas with a temperature range comparable to observations, in all the incoming models. Simulations suggest Mach numbers of roughly 1--2.

\item X-rays: All the scenarios exhibit detectable X-ray excess in the intermediate region between the two clusters. The best simulations overestimate the amplitude of the X-ray excess by a factor of $\sim$2--3.

\item Best scenarios: The model of tangential approach (B) proved to be the one that most closely matches the features of the X-ray surface brightness profiles. Models of radial (A) and vertical (C) approach were shown to be marginally compatible as well.

\item Disfavoured scenarios: The outgoing scenario (E) is generally disfavoured for failing to meet observational constraints, as well as the post-apocentric approach~(D).

\end{itemize}

New spectroscopic data on A1758S should enable a refined modelling taking into account the internal structure of A1758S itself. Simultaneously, future refinements of these simulations should aim to introduce a line-of-sight velocity component into the current best models to evaluate how they are impacted. Future observations with the next-generation X-ray observatory \textit{Athena} \citep{Nandra2013} might probe the bridge region more deeply. If A1758N and A1758S are to merge in the future, as seems likely from the simulations presented here, then the resulting object will be among the most massive known in the Universe.

\begin{acknowledgments}

\section*{acknowledgments}

We thank the anonymous referee for the helpful suggestions which improved the results of the paper. The authors acknowledge the National Laboratory for Scientific Computing (LNCC/MCTI, Brazil) for providing HPC resources of the SDumont supercomputer, which have contributed to the research results reported within this paper. REGM acknowledges support from the Brazilian agency \textit{Conselho Nacional de Desenvolvimento Cient\'ifico e Tecnol\'ogico} (CNPq) through grants 303426/2018-7 and 307205/2021-5. REGM acknowledges support from \textit{Funda\c c\~ao de Apoio \`a Ci\^encia, Tecnologia e Inova\c c\~ao do Paran\'a} through grant 18.148.096-3 -- NAPI \textit{Fen\^omenos Extremos do Universo}. RPA acknowledges the financial support from UTFPR. GBLN acknowledges support from CNPq through grant 303130/2019-9.

\end{acknowledgments}

\bibliography{paper.bib}{}

\begin{thebibliography}{}
\expandafter\ifx\csname natexlab\endcsname\relax\def\natexlab#1{#1}\fi
\providecommand{\url}[1]{\href{#1}{#1}}
\providecommand{\dodoi}[1]{doi:~\href{http://doi.org/#1}{\nolinkurl{#1}}}
\providecommand{\doeprint}[1]{\href{http://ascl.net/#1}{\nolinkurl{http://ascl.net/#1}}}
\providecommand{\doarXiv}[1]{\href{https://arxiv.org/abs/#1}{\nolinkurl{https://arxiv.org/abs/#1}}}

\bibitem[{{Akamatsu} {et~al.}(2016){Akamatsu}, {Gu}, {Shimwell}, {Mernier},
  {Mao}, {Urdampilleta}, {de Plaa}, {R{\"o}ttgering}, \&
  {Kaastra}}]{Akamatsu2016}
{Akamatsu}, H., {Gu}, L., {Shimwell}, T.~W., {et~al.} 2016, \aap, 593, L7,
  \dodoi{10.1051/0004-6361/201629275}

\bibitem[{{Akamatsu} {et~al.}(2017){Akamatsu}, {Fujita}, {Akahori}, {Ishisaki},
  {Hayashida}, {Hoshino}, {Mernier}, {Yoshikawa}, {Sato}, \&
  {Kaastra}}]{Akamatsu2017}
{Akamatsu}, H., {Fujita}, Y., {Akahori}, T., {et~al.} 2017, \aap, 606, A1,
  \dodoi{10.1051/0004-6361/201730497}

\bibitem[{{Albuquerque} {et~al.}(2024){Albuquerque}, {Machado}, \&
  {Monteiro-Oliveira}}]{Albuquerque2024}
{Albuquerque}, R.~P., {Machado}, R. E.~G., \& {Monteiro-Oliveira}, R. 2024,
  \mnras, 530, 2146, \dodoi{10.1093/mnras/stae1004}

\bibitem[{{Andrade-Santos} {et~al.}(2015){Andrade-Santos}, {Jones}, {Forman},
  {Murray}, {Kraft}, {Vikhlinin}, {van Weeren}, {Nulsen}, {David}, {Dawson},
  {Arnaud}, {Pointecouteau}, {Pratt}, \& {Melin}}]{AndradeSantos2015}
{Andrade-Santos}, F., {Jones}, C., {Forman}, W.~R., {et~al.} 2015, \apj, 803,
  108, \dodoi{10.1088/0004-637X/803/2/108}

\bibitem[{{Bianconi} {et~al.}(2020){Bianconi}, {Smith}, {Haines}, {McGee},
  {Finoguenov}, \& {Egami}}]{Bianconi2020}
{Bianconi}, M., {Smith}, G.~P., {Haines}, C.~P., {et~al.} 2020, \mnras, 492,
  4599, \dodoi{10.1093/mnras/staa085}

\bibitem[{{Biffi} {et~al.}(2013){Biffi}, {Dolag}, \&
  {B{\"o}hringer}}]{Biffi2013}
{Biffi}, V., {Dolag}, K., \& {B{\"o}hringer}, H. 2013, \mnras, 428, 1395,
  \dodoi{10.1093/mnras/sts120}

\bibitem[{{Biffi} {et~al.}(2012){Biffi}, {Dolag}, {B{\"o}hringer}, \&
  {Lemson}}]{Biffi2012}
{Biffi}, V., {Dolag}, K., {B{\"o}hringer}, H., \& {Lemson}, G. 2012, \mnras,
  420, 3545, \dodoi{10.1111/j.1365-2966.2011.20278.x}

\bibitem[{{Biffi} {et~al.}(2022){Biffi}, {Dolag}, {Reiprich}, {Veronica},
  {Ramos-Ceja}, {Bulbul}, {Ota}, \& {Ghirardini}}]{Biffi2022}
{Biffi}, V., {Dolag}, K., {Reiprich}, T.~H., {et~al.} 2022, \aap, 661, A17,
  \dodoi{10.1051/0004-6361/202141107}

\bibitem[{{Boschin} {et~al.}(2012){Boschin}, {Girardi}, {Barrena}, \&
  {Nonino}}]{Boschin2012}
{Boschin}, W., {Girardi}, M., {Barrena}, R., \& {Nonino}, M. 2012, \aap, 540,
  A43, \dodoi{10.1051/0004-6361/201118076}

\bibitem[{{Botteon} {et~al.}(2018){Botteon}, {Shimwell}, {Bonafede},
  {Dallacasa}, {Brunetti}, {Mandal}, {van Weeren}, {Br{\"u}ggen}, {Cassano},
  {de Gasperin}, {Hoang}, {Hoeft}, {R{\"o}ttgering}, {Savini}, {White},
  {Wilber}, \& {Venturi}}]{Botteon2018}
{Botteon}, A., {Shimwell}, T.~W., {Bonafede}, A., {et~al.} 2018, \mnras, 478,
  885, \dodoi{10.1093/mnras/sty1102}

\bibitem[{{Botteon} {et~al.}(2020){Botteon}, {van Weeren}, {Brunetti}, {de
  Gasperin}, {Intema}, {Osinga}, {Di Gennaro}, {Shimwell}, {Bonafede},
  {Br{\"u}ggen}, {Cassano}, {Cuciti}, {Dallacasa}, {Gastaldello}, {Mandal},
  {Rossetti}, \& {R{\"o}ttgering}}]{Botteon2020}
{Botteon}, A., {van Weeren}, R.~J., {Brunetti}, G., {et~al.} 2020, \mnras, 499,
  L11, \dodoi{10.1093/mnrasl/slaa142}

\bibitem[{{Brunetti} \& {Vazza}(2020)}]{Brunetti2020}
{Brunetti}, G., \& {Vazza}, F. 2020, \prl, 124, 051101,
  \dodoi{10.1103/PhysRevLett.124.051101}

\bibitem[{{Chadayammuri} {et~al.}(2022){Chadayammuri}, {ZuHone}, {Nulsen},
  {Nagai}, {Felix}, {Andrade-Santos}, {King}, \& {Russell}}]{Chadayammuri22}
{Chadayammuri}, U., {ZuHone}, J., {Nulsen}, P., {et~al.} 2022, \mnras, 509,
  1201, \dodoi{10.1093/mnras/stab2629}

\bibitem[{{Clowe} {et~al.}(2006){Clowe}, {Brada{\v c}}, {Gonzalez},
  {Markevitch}, {Randall}, {Jones}, \& {Zaritsky}}]{Clowe2006}
{Clowe}, D., {Brada{\v c}}, M., {Gonzalez}, A.~H., {et~al.} 2006, \apjl, 648,
  L109, \dodoi{10.1086/508162}

\bibitem[{{Dahle} {et~al.}(2002){Dahle}, {Kaiser}, {Irgens}, {Lilje}, \&
  {Maddox}}]{Dahle2002}
{Dahle}, H., {Kaiser}, N., {Irgens}, R.~J., {Lilje}, P.~B., \& {Maddox}, S.~J.
  2002, \apjs, 139, 313, \dodoi{10.1086/338678}

\bibitem[{{David} \& {Kempner}(2004)}]{David2004}
{David}, L.~P., \& {Kempner}, J. 2004, \apj, 613, 831, \dodoi{10.1086/423195}

\bibitem[{{Dehnen}(1993)}]{Dehnen1993}
{Dehnen}, W. 1993, \mnras, 265, 250

\bibitem[{{Donnert} {et~al.}(2017){Donnert}, {Beck}, {Dolag}, \&
  {R{\"o}ttgering}}]{Donnert17}
{Donnert}, J.~M.~F., {Beck}, A.~M., {Dolag}, K., \& {R{\"o}ttgering}, H.~J.~A.
  2017, \mnras, 471, 4587, \dodoi{10.1093/mnras/stx1819}

\bibitem[{{Doubrawa} {et~al.}(2020){Doubrawa}, {Machado}, {Lagan{\'a}}, {Lima
  Neto}, {Monteiro-Oliveira}, \& {Cypriano}}]{Doubrawa2020}
{Doubrawa}, L., {Machado}, R.~E.~G., {Lagan{\'a}}, T.~F., {et~al.} 2020,
  \mnras, 495, 2022, \dodoi{10.1093/mnras/staa1051}

\bibitem[{{Durret} {et~al.}(2011){Durret}, {Lagan{\'a}}, \&
  {Haider}}]{Durret2011}
{Durret}, F., {Lagan{\'a}}, T.~F., \& {Haider}, M. 2011, \aap, 529, A38,
  \dodoi{10.1051/0004-6361/201015978}

\bibitem[{{Govoni} {et~al.}(2019){Govoni}, {Orr{\`u}}, {Bonafede}, {Iacobelli},
  {Paladino}, {Vazza}, {Murgia}, {Vacca}, {Giovannini}, {Feretti}, {Loi},
  {Bernardi}, {Ferrari}, {Pizzo}, {Gheller}, {Manti}, {Br{\"u}ggen},
  {Brunetti}, {Cassano}, {de Gasperin}, {En{\ss}lin}, {Hoeft}, {Horellou},
  {Junklewitz}, {R{\"o}ttgering}, {Scaife}, {Shimwell}, {van Weeren}, \&
  {Wise}}]{Govoni2019}
{Govoni}, F., {Orr{\`u}}, E., {Bonafede}, A., {et~al.} 2019, Science, 364, 981,
  \dodoi{10.1126/science.aat7500}

\bibitem[{{Ha} {et~al.}(2018){Ha}, {Ryu}, \& {Kang}}]{Ha2018}
{Ha}, J.-H., {Ryu}, D., \& {Kang}, H. 2018, \apj, 857, 26,
  \dodoi{10.3847/1538-4357/aab4a2}

\bibitem[{{Haines} {et~al.}(2013){Haines}, {Pereira}, {Smith}, {Egami},
  {Sanderson}, {Babul}, {Finoguenov}, {Merluzzi}, {Busarello}, {Rawle}, \&
  {Okabe}}]{Haines2013}
{Haines}, C.~P., {Pereira}, M.~J., {Smith}, G.~P., {et~al.} 2013, \apj, 775,
  126, \dodoi{10.1088/0004-637X/775/2/126}

\bibitem[{{Haines} {et~al.}(2018){Haines}, {Finoguenov}, {Smith}, {Babul},
  {Egami}, {Mazzotta}, {Okabe}, {Pereira}, {Bianconi}, {McGee}, {Ziparo},
  {Campusano}, \& {Loyola}}]{Haines2018}
{Haines}, C.~P., {Finoguenov}, A., {Smith}, G.~P., {et~al.} 2018, \mnras, 477,
  4931, \dodoi{10.1093/mnras/sty651}

\bibitem[{{Hernquist}(1990)}]{Hernquist1990}
{Hernquist}, L. 1990, \apj, 356, 359, \dodoi{10.1086/168845}

\bibitem[{{Landau} \& {Lifshitz}(1959)}]{Landau1959}
{Landau}, L.~D., \& {Lifshitz}, E.~M. 1959, {\textit{Fluid mechanics}}
  (Pergamon Press)

\bibitem[{{Lee} {et~al.}(2020){Lee}, {Jee}, {Kang}, {Ryu}, {Kimm}, \&
  {Br{\"u}ggen}}]{Lee2020}
{Lee}, W., {Jee}, M.~J., {Kang}, H., {et~al.} 2020, \apj, 894, 60,
  \dodoi{10.3847/1538-4357/ab855f}

\bibitem[{{Louren{\c{c}}o} {et~al.}(2020){Louren{\c{c}}o}, {Lopes},
  {Lagan{\'a}}, {Nascimento}, {Machado}, {Moura}, {Jaff{\'e}}, {Ribeiro},
  {Vulcani}, {Moretti}, \& {Riguccini}}]{Lourenco2020}
{Louren{\c{c}}o}, A. C.~C., {Lopes}, P.~A.~A., {Lagan{\'a}}, T.~F., {et~al.}
  2020, \mnras, 498, 835, \dodoi{10.1093/mnras/staa2464}

\bibitem[{{Machado} {et~al.}(2022){Machado}, {Lagan{\'a}}, {Souza}, {Caproni},
  {Antas}, \& {Mello-Terencio}}]{Machado2022}
{Machado}, R.~E.~G., {Lagan{\'a}}, T.~F., {Souza}, G.~S., {et~al.} 2022,
  \mnras, 515, 581, \dodoi{10.1093/mnras/stac1829}

\bibitem[{{Machado} \& {Lima Neto}(2013)}]{Machado2013}
{Machado}, R.~E.~G., \& {Lima Neto}, G.~B. 2013, \mnras, 430, 3249,
  \dodoi{10.1093/mnras/stt127}

\bibitem[{{Machado} \& {Lima Neto}(2015)}]{MachadoLimaNeto2015}
---. 2015, \mnras, 447, 2915, \dodoi{10.1093/mnras/stu2669}

\bibitem[{{Machado} {et~al.}(2015){Machado}, {Monteiro-Oliveira}, {Lima Neto},
  \& {Cypriano}}]{Machado2015}
{Machado}, R.~E.~G., {Monteiro-Oliveira}, R., {Lima Neto}, G.~B., \&
  {Cypriano}, E.~S. 2015, \mnras, 451, 3309, \dodoi{10.1093/mnras/stv1162}

\bibitem[{{Molnar} {et~al.}(2013){Molnar}, {Chiu}, {Broadhurst}, \&
  {Stadel}}]{Molnar2013}
{Molnar}, S.~M., {Chiu}, I. N.~T., {Broadhurst}, T., \& {Stadel}, J.~G. 2013,
  \apj, 779, 63, \dodoi{10.1088/0004-637X/779/1/63}

\bibitem[{{Monteiro-Oliveira}(2022)}]{Monteiro-Oliveira22}
{Monteiro-Oliveira}, R. 2022, \mnras, 515, 3674, \dodoi{10.1093/mnras/stac2053}

\bibitem[{{Monteiro-Oliveira} {et~al.}(2017){Monteiro-Oliveira}, {Cypriano},
  {Machado}, {Lima Neto}, {Ribeiro}, {Sodr{\'e}}, \& {Dupke}}]{Monteiro2017}
{Monteiro-Oliveira}, R., {Cypriano}, E.~S., {Machado}, R.~E.~G., {et~al.} 2017,
  \mnras, 466, 2614, \dodoi{10.1093/mnras/stw3238}

\bibitem[{{Moura} {et~al.}(2021){Moura}, {Machado}, \&
  {Monteiro-Oliveira}}]{Moura2021}
{Moura}, M.~T., {Machado}, R. E.~G., \& {Monteiro-Oliveira}, R. 2021, \mnras,
  500, 1858, \dodoi{10.1093/mnras/staa3399}

\bibitem[{{Nandra} {et~al.}(2013){Nandra}, {Barret}, {Barcons}, {Fabian}, {den
  Herder}, {Piro}, {Watson}, {Adami}, {Aird}, {Afonso}, {Alexander},
  {Argiroffi}, {Amati}, {Arnaud}, {Atteia}, {Audard}, {Badenes}, {Ballet},
  {Ballo}, {Bamba}, {Bhardwaj}, {Stefano Battistelli}, {Becker}, {De Becker},
  {Behar}, {Bianchi}, {Biffi}, {B{\^\i}rzan}, {Bocchino}, {Bogdanov}, {Boirin},
  {Boller}, {Borgani}, {Borm}, {Bouch{\'e}}, {Bourdin}, {Bower}, {Braito},
  {Branchini}, {Branduardi-Raymont}, {Bregman}, {Brenneman}, {Brightman},
  {Br{\"u}ggen}, {Buchner}, {Bulbul}, {Brusa}, {Bursa}, {Caccianiga},
  {Cackett}, {Campana}, {Cappelluti}, {Cappi}, {Carrera}, {Ceballos},
  {Christensen}, {Chu}, {Churazov}, {Clerc}, {Corbel}, {Corral}, {Comastri},
  {Costantini}, {Croston}, {Dadina}, {D'Ai}, {Decourchelle}, {Della Ceca},
  {Dennerl}, {Dolag}, {Done}, {Dovciak}, {Drake}, {Eckert}, {Edge}, {Ettori},
  {Ezoe}, {Feigelson}, {Fender}, {Feruglio}, {Finoguenov}, {Fiore}, {Galeazzi},
  {Gallagher}, {Gandhi}, {Gaspari}, {Gastaldello}, {Georgakakis},
  {Georgantopoulos}, {Gilfanov}, {Gitti}, {Gladstone}, {Goosmann}, {Gosset},
  {Grosso}, {Guedel}, {Guerrero}, {Haberl}, {Hardcastle}, {Heinz}, {Alonso
  Herrero}, {Herv{\'e}}, {Holmstrom}, {Iwasawa}, {Jonker}, {Kaastra}, {Kara},
  {Karas}, {Kastner}, {King}, {Kosenko}, {Koutroumpa}, {Kraft}, {Kreykenbohm},
  {Lallement}, {Lanzuisi}, {Lee}, {Lemoine-Goumard}, {Lobban}, {Lodato},
  {Lovisari}, {Lotti}, {McCharthy}, {McNamara}, {Maggio}, {Maiolino}, {De
  Marco}, {de Martino}, {Mateos}, {Matt}, {Maughan}, {Mazzotta}, {Mendez},
  {Merloni}, {Micela}, {Miceli}, {Mignani}, {Miller}, {Miniutti}, {Molendi},
  {Montez}, {Moretti}, {Motch}, {Naz{\'e}}, {Nevalainen}, {Nicastro}, {Nulsen},
  {Ohashi}, {O'Brien}, {Osborne}, {Oskinova}, {Pacaud}, {Paerels}, {Page},
  {Papadakis}, {Pareschi}, {Petre}, {Petrucci}, {Piconcelli}, {Pillitteri},
  {Pinto}, {de Plaa}, {Pointecouteau}, {Ponman}, {Ponti}, {Porquet}, {Pounds},
  {Pratt}, {Predehl}, {Proga}, {Psaltis}, {Rafferty}, {Ramos-Ceja}, {Ranalli},
  {Rasia}, {Rau}, {Rauw}, {Rea}, {Read}, {Reeves}, {Reiprich}, {Renaud},
  {Reynolds}, {Risaliti}, {Rodriguez}, {Rodriguez Hidalgo}, {Roncarelli},
  {Rosario}, {Rossetti}, {Rozanska}, {Rovilos}, {Salvaterra}, {Salvato}, {Di
  Salvo}, {Sanders}, {Sanz-Forcada}, {Schawinski}, {Schaye}, {Schwope},
  {Sciortino}, {Severgnini}, {Shankar}, {Sijacki}, {Sim}, {Schmid}, {Smith},
  {Steiner}, {Stelzer}, {Stewart}, {Strohmayer}, {Str{\"u}der}, {Sun}, {Takei},
  {Tatischeff}, {Tiengo}, {Tombesi}, {Trinchieri}, {Tsuru}, {Ud-Doula},
  {Ursino}, {Valencic}, {Vanzella}, {Vaughan}, {Vignali}, {Vink}, {Vito},
  {Volonteri}, {Wang}, {Webb}, {Willingale}, {Wilms}, {Wise}, {Worrall},
  {Young}, {Zampieri}, {In't Zand}, {Zane}, {Zezas}, {Zhang}, \&
  {Zhuravleva}}]{Nandra2013}
{Nandra}, K., {Barret}, D., {Barcons}, X., {et~al.} 2013, arXiv e-prints,
  arXiv:1306.2307, \dodoi{10.48550/arXiv.1306.2307}

\bibitem[{{Okabe} \& {Umetsu}(2008)}]{Okabe2008}
{Okabe}, N., \& {Umetsu}, K. 2008, \pasj, 60, 345,
  \dodoi{10.1093/pasj/60.2.345}

\bibitem[{{Paterno-Mahler} {et~al.}(2014){Paterno-Mahler}, {Randall}, {Bulbul},
  {Andrade-Santos}, {Blanton}, {Jones}, {Murray}, \&
  {Johnson}}]{PaternoMahler2014}
{Paterno-Mahler}, R., {Randall}, S.~W., {Bulbul}, E., {et~al.} 2014, \apj, 791,
  104, \dodoi{10.1088/0004-637X/791/2/104}

\bibitem[{{Ragozzine} {et~al.}(2012){Ragozzine}, {Clowe}, {Markevitch},
  {Gonzalez}, \& {Brada{\v c}}}]{Ragozzine2012}
{Ragozzine}, B., {Clowe}, D., {Markevitch}, M., {Gonzalez}, A.~H., \& {Brada{\v
  c}}, M. 2012, \apj, 744, 94, \dodoi{10.1088/0004-637X/744/2/94}

\bibitem[{{Ricker} \& {Sarazin}(2001)}]{Ricker2001}
{Ricker}, P.~M., \& {Sarazin}, C.~L. 2001, \apj, 561, 621,
  \dodoi{10.1086/323365}

\bibitem[{{Ritchie} \& {Thomas}(2002)}]{Ritchie2002}
{Ritchie}, B.~W., \& {Thomas}, P.~A. 2002, \mnras, 329, 675,
  \dodoi{10.1046/j.1365-8711.2002.05027.x}

\bibitem[{{Rizza} {et~al.}(1998){Rizza}, {Burns}, {Ledlow}, {Owen}, {Voges}, \&
  {Bliton}}]{Rizza1998}
{Rizza}, E., {Burns}, J.~O., {Ledlow}, M.~J., {et~al.} 1998, \mnras, 301, 328,
  \dodoi{10.1046/j.1365-8711.1998.01972.x}

\bibitem[{{Roediger} {et~al.}(2011){Roediger}, {Br{\"u}ggen}, {Simionescu},
  {B{\"o}hringer}, {Churazov}, \& {Forman}}]{Roediger2011}
{Roediger}, E., {Br{\"u}ggen}, M., {Simionescu}, A., {et~al.} 2011, \mnras,
  413, 2057, \dodoi{10.1111/j.1365-2966.2011.18279.x}

\bibitem[{{Roettiger} {et~al.}(1997){Roettiger}, {Loken}, \&
  {Burns}}]{Roettiger1997}
{Roettiger}, K., {Loken}, C., \& {Burns}, J.~O. 1997, \apjs, 109, 307,
  \dodoi{10.1086/312979}

\bibitem[{{Ruggiero} \& {Lima Neto}(2017)}]{Ruggiero2017}
{Ruggiero}, R., \& {Lima Neto}, G.~B. 2017, \mnras, 468, 4107,
  \dodoi{10.1093/mnras/stx744}

\bibitem[{{Schellenberger} {et~al.}(2019){Schellenberger}, {David},
  {O'Sullivan}, {Vrtilek}, \& {Haines}}]{Schellenberger2019}
{Schellenberger}, G., {David}, L., {O'Sullivan}, E., {Vrtilek}, J.~M., \&
  {Haines}, C.~P. 2019, \apj, 882, 59, \dodoi{10.3847/1538-4357/ab35e4}

\bibitem[{{Springel}(2005)}]{Springel2005}
{Springel}, V. 2005, \mnras, 364, 1105,
  \dodoi{10.1111/j.1365-2966.2005.09655.x}

\bibitem[{{Springel} \& {Farrar}(2007)}]{Springel2007}
{Springel}, V., \& {Farrar}, G.~R. 2007, \mnras, 380, 911,
  \dodoi{10.1111/j.1365-2966.2007.12159.x}

\bibitem[{{Takizawa}(1999)}]{Takizawa1999}
{Takizawa}, M. 1999, \apj, 520, 514, \dodoi{10.1086/307497}

\bibitem[{{Tittley} \& {Henriksen}(2001)}]{Tittley2001}
{Tittley}, E.~R., \& {Henriksen}, M. 2001, \apj, 563, 673,
  \dodoi{10.1086/323955}

\bibitem[{{Ursino} {et~al.}(2015){Ursino}, {Galeazzi}, {Gupta}, {Kelley},
  {Mitsuishi}, {Ohashi}, \& {Sato}}]{Ursino2015}
{Ursino}, E., {Galeazzi}, M., {Gupta}, A., {et~al.} 2015, \apj, 806, 211,
  \dodoi{10.1088/0004-637X/806/2/211}

\bibitem[{{van Weeren} {et~al.}(2011){van Weeren}, {Br{\"u}ggen},
  {R{\"o}ttgering}, \& {Hoeft}}]{vanWeeren2011}
{van Weeren}, R.~J., {Br{\"u}ggen}, M., {R{\"o}ttgering}, H.~J.~A., \& {Hoeft},
  M. 2011, \mnras, 418, 230, \dodoi{10.1111/j.1365-2966.2011.19478.x}

\bibitem[{{van Weeren} {et~al.}(2019){van Weeren}, {de Gasperin}, {Akamatsu},
  {Br{\"u}ggen}, {Feretti}, {Kang}, {Stroe}, \& {Zandanel}}]{vanWeeren2019}
{van Weeren}, R.~J., {de Gasperin}, F., {Akamatsu}, H., {et~al.} 2019, \ssr,
  215, 16, \dodoi{10.1007/s11214-019-0584-z}

\bibitem[{{Walker} {et~al.}(2017){Walker}, {Hlavacek-Larrondo},
  {Gendron-Marsolais}, {Fabian}, {Intema}, {Sanders}, {Bamford}, \& {van
  Weeren}}]{Walker2017}
{Walker}, S.~A., {Hlavacek-Larrondo}, J., {Gendron-Marsolais}, M., {et~al.}
  2017, \mnras, 468, 2506, \dodoi{10.1093/mnras/stx640}

\bibitem[{{Wegner} {et~al.}(2017){Wegner}, {Umetsu}, {Molnar}, {Nonino},
  {Medezinski}, {Andrade-Santos}, {Bogdan}, {Lovisari}, {Forman}, \&
  {Jones}}]{Wegner2017}
{Wegner}, G.~A., {Umetsu}, K., {Molnar}, S.~M., {et~al.} 2017, \apj, 844, 67,
  \dodoi{10.3847/1538-4357/aa784a}

\bibitem[{{Werner} {et~al.}(2008){Werner}, {Finoguenov}, {Kaastra},
  {Simionescu}, {Dietrich}, {Vink}, \& {B{\"o}hringer}}]{Werner2008}
{Werner}, N., {Finoguenov}, A., {Kaastra}, J.~S., {et~al.} 2008, \aap, 482,
  L29, \dodoi{10.1051/0004-6361:200809599}

\bibitem[{{ZuHone} \& {Hallman}(2016)}]{ZuHone2016}
{ZuHone}, J.~A., \& {Hallman}, E.~J. 2016, {pyXSIM: Synthetic X-ray
  observations generator}, Astrophysics Source Code Library, record
  ascl:1608.002.
\newblock \doeprint{1608.002}

\end{thebibliography}
\bibliographystyle{aasjournal}

\end{document}